\documentclass[prl, twocolumn]{revtex4}
\usepackage{amssymb}
\usepackage{amsmath}
\usepackage{epsfig}
\usepackage{bm}

\begin{document}

\title{When will we have a quantum computer?}

\author{M.I. Dyakonov}

\affiliation{Laboratoire Charles Coulomb, Universit\'e Montpellier, CNRS, France}

\maketitle

{\bf 1. Introduction, historical background}\smallskip 

The idea of quantum computing was first put forward in a rather
vague form by the Russian mathematician Yuri Manin in 1980. In 1981,
it was independently proposed by Richard Feynman. Realizing that
(because of the exponential increase of the number of quantum states)
computer simulations of quantum systems become impossible when the
system is large enough, he advanced the idea that to make them efficient 
the computer itself should operate in the quantum mode: ``Nature
isn\textsc{\char13}t classical and if you want to make a simulation of Nature, 
you\textsc{\char13}d better make it quantum mechanical, and by golly 
it\textsc{\char13}s a wonderful problem, because it doesn\textsc{\char13}t 
look so easy''. In 1985, David Deutsch formally described the universal quantum 
computer, as a quantum analogue of the universal Turing machine.

The subject did not attract much attention until Peter Shor in 1994 
proposed an algorithm that could factor very large numbers on an {\it ideal}
quantum computer much faster compared to the conventional (classical) 
computer. This outstanding theoretical result has triggered an
explosion of general interest in quantum computing and many thousands of 
research papers, mostly theoretical, have been and still continue to be 
published at an increasing rate.

During the last 20 years one can hardly find an issue of any science
digest magazine, or even of a serious physical journal, that does not
address quantum computing. Quantum Information Centers are
opening all over the globe, funds are generously distributed, and
breathtaking perspectives are presented to the layman by enthusiastic
scientists and journalists. Many researchers feel obliged to justify
whatever research they are doing by claiming that it has some relevance to 
quantum computing. 

Computer scientists are proving and publishing new theorems related to quantum computers 
at a rate of {\it one article per day}. A huge number of proposals has been published for
various physical objects that could serve as quantum bits, or qubits. As of September 25, 
2018, Google gives 71,400,000 results for ``quantum computing'', and 331,000 results for 
``quantum computing with'', and these numbers increase every day. The impression has been 
created that quantum computing - this modern version of the Holy Grail - is going to be the 
next technological revolution of the 21st century.

When will we have useful quantum computers? The most optimistic
experts say: ``in 10 years''; others predict 20 to 30 years (note that those
expectations have remained unchanged during the last 20 years), and
the most cautious ones say: ``not in my lifetime''. The present author
belongs to the meager minority that has been answering ``not in any
foreseeable future''\cite{Dyakonov}, and this point of view is explained below.

{\bf The idea of quantum computing} is to store and process information in a way that is very 
different from that used in conventional computers, which basically operate with an assembly 
of on/off switches, physically realized as tiny transistors.

At a given moment the state of the {\it classical} computer is described by
a sequence ($\uparrow \downarrow \uparrow \uparrow \downarrow \uparrow \downarrow 
\downarrow$...), where $\uparrow$ and $\downarrow$ represent {\it bits} of information 
realized as the {\it on} and {\it off} states of individual transistors. With {\it N} 
transistors, there are $2^N$ different possible states of the computer. The
computation process consists in a sequence of switching some transistors between 
their $\uparrow$ and $\downarrow$ states according to a prescribed program.

In {\it quantum} computing one replaces the classical two-state element
by a quantum element with two {\it basic} states, known as the quantum bit,
or {\it qubit}. The simplest object of this kind is the electron internal angular
momentum, spin, with the peculiar quantum property of having only
two possible projections on {\it any} axis: +1/2 or $-$1/2 (in units of the
Planck constant $\hbar$). For some chosen axis, we can again denote the two
basic quantum states of the spin as $\uparrow$ and $\downarrow$.

However, an {\it arbitrary} spin state is described by the wave function
$\psi = a$$\uparrow$+ b$\downarrow$, where {\it a} and {\it b} are complex numbers, 
satisfying the normalization condition $|a|^2 + |b|^2 = 1$, so that $|a|^2$ and $|b|^2$ 
are the {\it probabilities} for the spin to be in the basic states $\uparrow$ and 
$\downarrow$ respectively.

Unlike the classical bit, that can be only in {\it one} of the two states,
$\uparrow$ or $\downarrow$, the qubit can be in a {\it continuum} of states defined 
by the quantum amplitudes {\it a} and {\it b}. Thus, in contrast to the classical bit, 
{\it\bf the qubit is a continuous object}.

This property is often described by the rather mystical and frightening statement that 
the qubit can exist {\it simultaneously} in both of its $\uparrow$ and $\downarrow$ states. 
(This is like saying that a vector in the {\it x-y} plane directed at 45 degrees to 
the {\it x}-axis simultaneously points both in the {\it x-} and {\it y-}directions -- a statement 
that is true in some sense, but does not have much useful content.)

Note that since {\it a} and {\it b} are complex numbers satisfying the normalization condition, 
and since the overall phase of the wave function is irrelevant, there remain two free parameters 
defining the state of a single qubit ({\bf exactly like for a classical vector} whose orientation 
in space is defined by two polar angles). This analogy does not apply any longer when the number 
of qubits is 2 or more.

With two qubits, one has $2^2$ basic states: $\uparrow\uparrow, \uparrow\downarrow,  
 \downarrow\uparrow$, and $\downarrow\downarrow$.  Accordingly, they are described by the 
wave function $\psi=a\uparrow\uparrow+b\uparrow\downarrow+c\downarrow\uparrow+d\downarrow\downarrow$ 
with 4 complex amplitudes {\it a, b, c}, and {\it d}. In the general case of {\it N} qubits, the state 
of the system is described by $2^N$ complex amplitudes restricted by the normalization condition only.

{\bf While the state of the classical computer with $N$ bits at any given moment coincides with {\it one} 
of its $2^N$ possible discreet states, the state of a quantum computer with $N$ qubits is described by the 
values of $2^N$ {\it continuous} variables, the quantum amplitudes}. 

This is at the origin of the supposed power of the quantum computer, but it is also the reason for 
it\textsc{\char13}s great fragility and vulnerability. The information processing is supposed to be done 
by applying unitary transformations (quantum gates), that change these amplitudes
{\it a, b, c...} in a precise and controlled manner.

The number of qubits needed to have a useful machine (i.e. one that can compete with your laptop 
in solving certain problems, such as factoring very large numbers by Shor\textsc{\char13}s algorithm) is 
estimated to be $10^3-10^5$ . As a result, the number of continuous variables describing the state of such a 
quantum computer at any given moment is at least $2^{1000} \sim 10^{300}$) which is much, much greater 
than the number of particles in the whole Universe (only $\sim 10^{80}$)!

At this point a normal engineer, or an experimentalist, loses interest. Indeed, possible errors in 
a classical computer consist in the fact that one or more transistors are switched off instead of 
being switched on, or vice versa. This certainly is an unwanted occurrence, but can be dealt
with by relatively simple methods employing {\it redundance}.

In contrast, accomplishing the Sisyphean task of keeping under control $10^{300}$ continuous 
variables is absolutely unimaginable. However, the QC theorists have succeeded in transmitting 
to the media and to the general public the belief that the feasibility of large-scale quantum
computing has been {\it proven} via the famous threshold theorem: once the error per qubit per 
gate is below a certain value, indefinitely long quantum computation becomes feasible, at a cost 
of substantially increasing the number of qubits (the logical qubit is encoded by several 
physical qubits). 

Very luckily, the number of qubits increases {\it only polynomially} with the size of computation, 
so that the total number of qubits needed must increase from $N = 10^3$ to $N = 10^6-10^9$ only 
(with a corresponding increase of the atrocious number of $2^N$ continuous parameters defining the 
state of the whole machine!) \cite{Lidar}.

In this context, Leonid Levin, professor of mathematics at Boston University, has made the 
following pertinent remark: {\it What thought experiments can probe the QC to be in the state described 
with the accuracy needed? I would allow to use the resources of the entire Universe, but not more!}

{\bf 2. Expert panels: 2018 {\it vs} 2002}\smallskip 

Seventeen years ago, in 2002, at the request of the Advanced Research and Development Activity (ARDA) 
agency of the United States government, a team of distinguished experts in quantum information established 
a roadmap \cite{arda} for quantum computing, with the following five- and ten-year goals:
 \\- encode a single qubit into the state of a logical qubit formed from several physical qubits;
 \\- perform repetitive error correction of the logical qubit; and
 \\- transfer the state of the logical qubit into the state of another set of physical qubits with high fidelity;
 \\- and by the year 2012, to implement a concatenated \cite{concat} error-correcting code.

The 2007 goal requires ``something on the order of ten physical qubits and multiple logic operations between them'', 
while the 2012 goal ``requires on the order of 50 physical qubits, exercises multiple logical qubits through the 
full range of operations required for fault-tolerant QC in order to perform a simple instance of a relevant quantum
algorithm''.

While a benevolent jury could consider the first two of the 2007 goals to be partly achieved by now, the 
expectations for the third 2007 goal, and especially for the 2012 goal, are {\bf wildly off the mark}. So are
some other predictions of the ARDA panel: ``As larger-scale quantum computers are developed over the next five and 
ten years, quantum simulation is likely to continue to be the application for which quantum computers can give 
substantial improvements over classical computation''.

Very recently, in late 2018, another expert panel assembled by the U.S. National Academies of Science, 
Engineering and Medicine issued a detailed 205-page report discussing some of the challenges facing QC as
a technology of practical value \cite{National}. The authors of the report state that {\it no quantum computer} 
will be capable of breaking cryptographic codes based prime number factoring within the next decade, and do 
not provide any opinion on whether or not this will be possible in a more distant future.  

{\bf Experimental studies} related to the idea of quantum computing make only a tiny part of the huge 
QC literature. They represent the {\it nec plus ultra} of the modern experimental technique, 
they are extremely difficult and inspire respect and admiration. The goal of such proof-of-principle 
experiments is to show the possibility to realize the basic quantum operations, as well as to demonstrate 
some elements of quantum algorithms. The number of qubits used is below 10, usually from 3 to 5.

Apparently, going from 5 qubits to 50 (the goal set by the ARDA Experts Panel roadmap for the year 2012!) 
presents hardly surmountable experimental difficulties and the reasons for this should be understood. 
Most probably, they are related to the simple fact that $2^5$ = 32, while $2^{50}$ = 1,125,899,906,842,624.

By contrast, the {\bf theory} of quantum computing, which largely dominates the literature, does not appear 
to encounter any substantial difficulties in dealing with millions of qubits. Various noise models are
being considered, and it has been proved (under certain assumptions) that errors generated by ``local'' 
noise can be corrected by carefully designed and very ingenious methods, involving, among other tricks,
massive parallelism: many thousands of gates should be applied simultaneously to different pairs of qubits 
and many thousands of measurements should be done simultaneously too.

The ARDA Experts Panel also claimed: ``It has been established, under certain assumptions, that if a 
threshold precision per gate operation could be achieved, quantum error correction would allow a
quantum computer to compute indefinitely''. Here, the key words are ``under certain assumptions'', 
however the distinguished experts did not address the crucial point of whether these assumptions can 
be realized in the physical world.

I argue that they can\textsc{\char13}t. In the physical world, continuous quantities (be they voltages 
or the parameters defining quantum-mechanical wave functions) can neither be measured nor manipulated 
exactly. To a mathematician, this might sound absurd, but this is the unquestionable reality of the world 
we live in. Sure, discrete quantities, like the number of students in a classroom or the number of transistors 
in the ``on'' state, {\it can} be known exactly. And {\it this} makes the great difference between a 
classical computer and the hypothetical quantum computer.

Indeed, all of the assumptions that theorists make about the preparation of qubits into a given 
state, the operation of the quantum gates, the reliability of the measurements, and so forth, 
cannot be fulfilled exactly. They can only be approached with some limited precision.
So, the question is: What precision is required? With what exactitude must, say, the $\sqrt{2}$ 
(an irrational number that enters into many of the relevant quantum operations) be experimentally 
realized? Can it be approximated as 1.41 or as 1.41421356237? There are no clear answers to these 
and many similar crucial questions.

An {\bf extremely important issue} is related to the energies of the $\uparrow$ and $\downarrow$ 
states. While the notion of {\it energy} is of primordial importance in all domains of physics, 
both classical and quantum, quite amazingly, it is not in the vocabulary of QC theorists. They 
implicitly assume that the energies of all $2^N$ states of an ensemble of qubits are {\it exactly 
equal}. Otherwise, the existence of an energy difference $\Delta E$ leads to oscillations of the 
quantum amplitudes with a frequency $\Omega = \Delta E/\hbar$, where $\hbar$ is the Planck constant, 
and this is a basic fact of quantum mechanics. (For example, one of the popular candidates
for a qubit, the electron spin, will make a precession around the direction of the Earth\textsc{\char13}s 
magnetic field with a frequency of $\sim$ 1 MHz. Should the Earth\textsc{\char13}s magnetic field 
be shielded, and if yes, with what precision?) 

Whatever is the nature of qubits, some energy differences will necessarily exist because of stray 
fields, various interactions, etc. resulting in a chaotic dynamics of the whole system, which will 
completely disorganize the performance of the quantum machine. I am not aware of any studies of this 
very general problem.

The problem of the accuracy required arises already at the first step, the preparation of the initial 
state of the quantum computer, which should be ($\uparrow \uparrow \uparrow ...$), or in conventional 
notation $|00000... >$, e.g. we start with all spins aligned in the {\it z}-direction, which will be the 
first task for the ``future quantum engineer''. However, where is the {\it z}-direction? Certainly, it 
can be defined arbitrarily, but only within a certain precision (like any continuous parameter). Aligning 
spins along this direction can also be done only approximately. So, instead of the desired $|00000... >$ 
state, inevitably we will have an admixture of all other states, hopefully with small amplitudes.

The same question (again, without any answers) concerns quantum gates, that is our manipulations with 
the qubits required to perform a meaningful quantum calculation. For example, the theorist proposes us
to flip the qubit, i.e. perform the operation $|0 > \rightarrow |1 >$ . Obviously, this again cannot be 
done exactly (especially, since the initial state $|0 >$ cannot be exact either), but the needed precision 
has not been established so far. 
\bigskip

{\bf 3. Quantum annealing} \smallskip

A completely different approach, initially started by the D-Wave company and now followed and developed 
by IBM, Google, Microsoft, and others, is based on using as qubits superconducting Josephson
junctions at ultra-low dilution fridge temperatures. Depending on some parameters of the system, Josephson 
junctions can operate either as classical two-state bits (and classical computers using Josephson logic
have been demonstrated), or as quantum bits.

This is not going to be the quantum computer everyone was talking about for the past 20 years, it will 
not be able to factor large numbers by Shor\textsc{\char13}s algorithm or to efficiently search databases 
by Grover\textsc{\char13}s quantum algorithm. Rather, it is supposed to perform ``quantum annealing''. After 
initial preparation, any system, whether classical or quantum, at low temperature will relax to its ground 
state. Calculating the ground state of more or less complex quantum systems, either analytically or numerically, 
is usually impossible and this is what originally inspired Feynman's vague idea of quantum computing.

Hence comes the idea of {\it simulating} a system of interacting qubits by an equivalent system of 
superconducting quantum circuits based on Josephson junctions. One does not do any quantum calculations 
by applying quantum gates, and quantum error correction is not needed either. One has just to measure 
the state of the system after annealing, more precisely, one can measure {\it some} of its $2^N$ parameters.

Such an approach is perfectly reasonable. However, Google claims that the 72-qubit superconducting chip 
in a 10-millikelvin dilution refrigerator (note that such a system is described by $2^{72} \sim 10^{21}$
quantum amplitudes) will prove that quantum computers can beat classical machines, and thus demonstrate 
``quantum supremacy''.

This claim appears to be somewhat exaggerated. The chip in question is not going to be a {\it quantum 
computer}, it will be only a specific quantum system (which might be quite interesting on its own) defined
by the way the Josephson junctions are interconnected. It is not entirely clear what will be the possible 
practical use of such systems. However, such modelling might provide some additional knowledge on the 
behavior of large and complicated quantum systems. 

Recently, a remarkable simulation of the Kosterlitz-Thouless phase transition was demonstrated in a 
network of Josephson superconducting rings arranged in a frustrated lattice \cite{King}.
\bigskip

{\bf 4. Conclusion} \smallskip
 
The hypothetical quantum computer is a system with an unimaginable number of continuous degrees of 
freedom - the values of the $2^N$ quantum amplitudes with $N\sim 10^3 - 10^5$ . These values 
{\it cannot be arbitrary}, they should be under our control with a high precision (which has yet to 
be defined).

In riding a bike, after some training, we learn to successfully control 3 degrees of freedom: 
the velocity, the direction, and the angle that our body makes with respect to the pavement. 
A circus artist manages to ride a one-wheel bike with 4 degrees of freedom. Now, imagine a bike
having 1000 (if not $2^{1000}$!) joints that allow free rotations of their parts with respect to 
each other. Will anybody be capable of riding this machine?

Thus, the answer to the question in title is: As soon as physicists and engineers learn to control 
this number of degrees of freedom, which means - never!
\medskip

{\it About the author}: Mikhail Dyakonov received the PhD (1966) in theoretical physics from Ioffe 
Physico-Technical Institute in Saint Petersburg (Leningrad), USSR. He worked at Ioffe Institute
until 1998 when he became professor at the University of Montpellier, France. He was elected an 
Honorary Member of Ioffe Institute in 2014. His fields of interest include physics of semiconductors, 
spin physics, and physics of 2D electrons. He is recipient of the State prize of USSR, Beller
lectureship award from the American Physical Society, and the Robin prize from the French Physical 
Society. His name is connected to several physical phenomena: Dyakonov-Perel spin relaxation mechanism, 
Dyakonov surface waves, Dyakonov-Shur instability. Together with V.I. Perel, he has predicted the Spin 
Hall Effect.

\end{document}